\documentclass[aps,prx,floatfix,superscriptaddress,noshowpacs,twocolumn]{revtex4-2}
\usepackage{graphics,graphicx,epsfig,amsmath,amssymb,epstopdf,wasysym,comment}
\usepackage{hyperref}
\usepackage{color}
\usepackage[normalem]{ulem}
\usepackage[dvipsnames]{xcolor}
\usepackage{braket,amsfonts}
\usepackage{dsfont}
\usepackage{hyperref}
\usepackage[dvipsnames]{xcolor}
\usepackage{amsthm}
\theoremstyle{remark}

\newcommand{\be}{\begin{equation}}
\newcommand{\ee}{\end{equation}}
\newcommand{\bea}{\begin{eqnarray}}
\newcommand{\eea}{\end{eqnarray}}
\newcommand{\ba}{\begin{eqnarray*}}
\newcommand{\ea}{\end{eqnarray*}}
\newcommand{\dagga}{{\phantom{\dagger}}}

\newcommand{\bk}{\mathbf{k}}

\newcommand{\bx}{\mathbf{x}}
\newcommand{\by}{\mathbf{y}}

\newcommand{\Ima}{\text{Im}}

\newcommand{\dis}{\displaystyle}

\newcommand{\fract}[2]{\frac{\dis #1}{\dis #2}}
\newcommand{\Tr}{\mathrm{Tr}}
\newcommand{\eqn}[1]{(\ref{#1})}

\renewcommand{\bra}[1]{\langle #1 \mid}
\renewcommand{\ket}[1]{\mid #1 \rangle}
\newenvironment{eqs}%
{\begin{equation} \begin{aligned}}%
{\end{aligned} \end{equation} }
\newcommand{\beal}{\begin{eqs}}
\newcommand{\eal}{\end{eqs}}
\newcommand{\bealn}{\beal\nonumber}
\newcommand{\bw}{\begin{widetext}}
\newcommand{\ew}{\end{widetext}}
\newcommand{\esp}[1]{\text{e}^{#1}}

\newcommand{\ep}{\epsilon}
\newcommand{\bd}[1]{{\boldsymbol{#1}}}
\renewcommand{\bf}[1]{{\mathbf{#1}}}%

\begin{document}
\title{Unified role of Green's function poles and zeros in correlated topological insulators}

\author{Andrea Blason}
\affiliation{International School for Advanced Studies (SISSA), Via Bonomea 265, I-34136 Trieste, Italy} 
\author{Michele Fabrizio}
\affiliation{International School for Advanced Studies (SISSA), Via Bonomea 265, I-34136 Trieste, Italy} 

\begin{abstract}
Green's function zeros, which can emerge only if correlation is strong, have been for long overlooked and believed to be devoid of any physical meaning, unlike Green's function poles. Here, we prove that Green's function zeros instead contribute on the same footing as poles 
to determine the topological character of an insulator. The key to the proof, worked out explicitly in 2D but easily extendable in 3D, 
is to express the topological invariant in terms of a \textit{quasiparticle} 
thermal Green's function matrix $G_*(i\ep,\bk)= 1/\big(i\ep-H_*(\ep,\bk)\big)$, with hermitian $H_*(\ep,\bk)$, by filtering out the positive definite \textit{quasiparticle} residue. In that way, the topological invariant is easily found to reduce to the TKNN formula for quasiparticles 
described by the non-interacting Hamiltonian $H_*(0,\bk)$. Since the poles 
of the quasiparticle Green's function $G_*(\ep,\bk)$ on the real frequency axis correspond to poles and zeros of the physical-particle Green's function $G(\ep,\bk)$, 
both of them equally determine the topological character of an insulator.  
\end{abstract}

                         
\maketitle

\section{Introduction}
The determinant of the retarded Green's function $G_0(\ep,\bk)$, with $\ep$ the frequency and $\bk$ the momentum, for periodic models of non-interacting electrons have poles whenever $\ep$ hits the dispersion energy $\ep_n(\bk)$ of a band $n$, i.e., a single-particle excitation. 
Similarly, the poles of the determinant of the fully-interacting $G(\ep,\bk)$ can be associated to coherent, i.e., with infinite lifetime, single-particle excitations, which thus have a clear physical meaning. The manifold in the Brillouin zone where these poles are at $\ep=0$ defines the Fermi surface, in which case the system is metallic. \\ 
However, the determinant of $G(\ep,\bk)$ in presence of interaction may also 
develop zeros~
\footnote{Equivalently, in the basis that diagonalises $G(\ep,\bk)$ for strictly real $\ep$, there may be elements along the diagonal that vanish at certain $\ep$ and $\bk$. That is evidently different from the case in which the Green's function, represented in a basis in which it is not diagonal, has diagonal terms that may cross zero. This circumstance has been discussed in \cite{PhysRevB.105.155112}. In particular, the diagonal elements of the non-interacting Green's function have only poles in the diagonal basis, while they may have also zeros in a generic non-diagonal basis, whose role has been analysed, e.g., in \cite{Slager-PRB2015,PhysRevResearch.4.023177,Obakpolor-PRB2022}},
whose manifold at $\ep=0$ defines the so-called Luttinger 
surface~\cite{Igor-PRB2003}. For long time, these zeros have not been given any physical significance, despite Volovik~\cite{Volovik-1} early on recognised that 
a Luttinger surface bears the same non-trivial topological content of a Fermi surface. Only recently, the Green's function zeros started to attract growing physical interest. For instance, it has been shown that a Luttinger surface defined by the simple roots of $\text{det}\big(G(0,\bk)\big)$ does sustain Landau's quasiparticles~\cite{mio-2}, 
even in non-symmetry breaking Mott insulators~\cite{mio-Mott}. Those quasiparticles have the same physical properties as conventional ones at a Fermi surface, 
with the major difference that they are incompressible~\cite{Jan} and do not contribute to charge transport~\cite{mio-Mott}. Elaborating on 
Volovik's observation~\cite{Volovik-1}, Gurarie~\cite{Gurarie-zeros-PRB2011} 
and Essin and Gurarie~\cite{Essing&Gurarie-PRB2011} have proposed that, upon increasing electron correlations, 
topological edge modes, i.e., edge poles of the Green's function,  
may transform into edges zeros without making the topological insulator a trivial one or  closing the single-particle gap. More recently, this intriguing scenario has been further explored in Ref.~\cite{Giorgio}, whose authors show that 
model topological insulators turn, upon rising interaction strength, into Mott insulators with topologically trivial lower and upper Hubbard bands, but with ingap valence and conduction bands of Green's function zeros that are topological and yield Green's function edge zeros. This result suggests that an edge-bulk correspondence exists also for Green's function zeros, thus clarifying the mechanism underlying the transformation of edge poles into edge zeros~\cite{Gurarie-zeros-PRB2011,Essing&Gurarie-PRB2011}.\\
All these small pieces of evidence suggest that Green's function zeros, that may arise only in strongly correlated systems, do have a physical meaning as important as that of Green's function poles. This connection has been uncovered when the zeros cross the chemical potential, thus in the presence of a Luttinger surface~\cite{Volovik-1,mio-2,mio-Mott}. However, despite the supporting evidences~\cite{Gurarie-zeros-PRB2011,Essing&Gurarie-PRB2011,Giorgio}, the direct role of Green's function zeros in assessing the topological character of an insulator has not been explicitly demonstrated. That is precisely the goal of the present work.

\section{TKNN formula for interacting insulators}
\label{TKNN formula for interacting insulators}

The expression of the topological invariant of two dimensional periodic insulators 
in presence of interaction, which coincides with the zero temperature Hall conductance in units of $e^2/2\pi\hbar$~\cite{Ishikawa-1, Ishikawa-2} 
at least when perturbation theory is valid, see Appendix \ref{Topological invariant in two dimensions}, reads  
\bw
\beal
W(G) &= \fract{1}{24\pi^2}\,\int d\ep\,d\bk\,\ep_{\mu\nu\rho}
\,\Tr\Big(G(i\ep,\bk)\;\partial_\mu G(i\ep,\bk)^{-1}\;
G(i\ep,\bk)\;\partial_\nu G(i\ep,\bk)^{-1}\;
G(i\ep,\bk)\;\partial_\rho G(i\ep,\bk)^{-1}\Big)\,,
\label{W}
\eal
\ew 
where $G(i\ep,\bk)=G(-i\ep,\bk)^\dagger$ is the interacting Green's functions in the Matsubara 
formalism, see Appendix~\ref{Properties of thermal Green's functions}, with $\ep$ the Matsubara frequency, and it is a matrix represented in a generic basis of 
single-particle wavefunctions. We hereafter assume that $G(i\ep,\bk)$ 
is invertible, which implies that the system is an insulator without any Luttinger surface. It can be easily demonstrated that the winding number $W(G_1 \, G_2) = W(G_1) + W(G_2) $ and that $W(G)=0$ if $G$ is hermitian, 
see Appendix \ref{Properties of the winding number}.\\

Zhong and Zhang have shown~\cite{Zhang-PRX2012} that the topological invariant \eqn{W} 
reduces to the TKNN expression of the quantised Hall conductance~\cite{TKNN} in which the role 
of the Bloch waves of the occupied bands is played by the eigenstates of the 
hermitian matrix $-G(0,\bk)^{-1}$ with negative eigenvalues. The proof is based on the observation that the two maps $(\ep,\bk)\to G(i\ep,\bk)^{-1}$ and $(\ep,\bk)\to i\ep+G(0,\bk)^{-1}$ are homotopic, and thus the winding number \eqn{W} of the former map 
coincides with that of the latter, which, in turns, reduces to the TKNN formula.\\

\noindent 
Here, we would like to prove explicitly the equivalence relation but using a 
different map $(\ep,\bk)\to G_*(i\ep,\bk)^{-1}$, which, we believe, has the more transparent physical meaning of the inverse of the \textit{quasiparticle} Green's function, which we discuss more extensively at the end of the section. Specifically, following~\cite{Jan,mio-Mott} we write the interacting Green's function matrix as, see also Appendix~\ref{Properties of thermal Green's functions},
\beal
G(i\ep,\bk) &= \fract{1}{\;i\ep-H_0(\bk)-\Sigma(i\ep,\bk)\;}\\
&=
\sqrt{Z(\ep,\bk)\;}\;\fract{1}{\;i\ep-H_*(\ep,\bk)\;}\;
\sqrt{Z(\ep,\bk)\;}\\
&\equiv \sqrt{Z(\ep,\bk)\;}\;G_*(i\ep,\bk)\;
\sqrt{Z(\ep,\bk)\;}\;,
\label{Green's}
\eal  
where $H_0(\bk)$ is the non-interacting Hamiltonian represented in the chosen basis of single-particle wavefunctions, 
\bealn
\Sigma(i\ep,\bk)&=\Sigma(-i\ep,\bk)^\dagger \equiv \Sigma_1(i\ep,\bk)+i\,\Sigma_2(i\ep,\bk)\\
\Sigma_1(i\ep,\bk) &= \Sigma_1(i\ep,\bk)^\dagger=\Sigma_1(-i\ep,\bk)\\
&=
\fract{\Sigma(i\ep,\bk)+\Sigma(-i\ep,\bk)}{2}\\
\Sigma_2(i\ep,\bk)&=\Sigma_2(i\ep,\bk)^\dagger=-\Sigma_2(-i\ep,\bk)\\
&=\fract{\Sigma(i\ep,\bk)-\Sigma(-i\ep,\bk)}{2i}
\eal
the self-energy matrix in that same basis, which accounts for all 
interaction effects, 
\beal
Z(\ep,\bk)&=Z(-\ep,\bk) = 
\Bigg(1-\fract{\;\Sigma_2(i\ep,\bk)\;}{\ep}\Bigg)^{-1}\,,
\label{Z}
\eal
a positive definite matrix if the system is insulating without a Luttinger 
surface, see Appendix~\ref{Properties of thermal Green's functions}, 
which can be regarded as the \textit{quasiparticle} residue, 
and 
\beal
&H_*(\ep,\bk)=H_*(-\ep,\bk)\\
& =
\sqrt{\,Z(\ep,\bk)\,}\,\Big(
H_0(\bk)+\Sigma_1(i\ep,\bk)\Big)\,\sqrt{\,Z(\ep,\bk)\,}\,,
\label{H*}
\eal
is the hermitian \textit{quasiparticle} Hamiltonian. It follows that the winding number \eqn{W} can be written as 
\bealn
W\big(G\big) &= W\Big(\sqrt{Z\,}\;G_*\,\sqrt{Z}\,\Big)=W\big(G_*\big) 
+W\big(Z\big)\\
& =W\big(G_*\big)\,,
\eal
since the winding number of the positive definite matrix $Z$ vanishes.\\ 
A further reason for choosing the map $(\ep,\bk)\to G_*(i\ep,\bk)^{-1}$ is that, under the analytic continuation on the real axis from above, $i\ep\to \ep+i0^+$, i.e., for the retarded components of Green's function and self-energy, the 
poles of $G_*(\ep,\bk)$ correspond to both poles and zeros of $G(\ep,\bk)$, i.e., 
the vanishing eigenvalues of the quasiparticle residue matrix \eqn{Z} on the real axis,  
thus making more explicit their deep connection. \\ 

\noindent
In the basis that diagonalises $H_*(\ep,\bk)$, i.e.,  
\bealn
H_*(\ep,\bk)\ket{\alpha(\ep,\bk)} &= \ep_\alpha(\ep,\bk)\,\ket{\alpha(\ep,\bk)}\,,
\eal
where one can choose $\ket{\alpha(\ep,\bk)}=\ket{\alpha(-\ep,\bk)}$, 
\beal
&W\big(G\big) = \fract{1}{24\pi^2}\,\int d\ep\,d\bk\,\ep_{\mu\nu\rho}\\
&\;\sum_{\alpha\beta\gamma}\,\fract{1}{\;i\ep-\ep_\alpha(\ep,\bk)\;}
\fract{1}{\;i\ep-\ep_\beta(\ep,\bk)\;}\fract{1}{\;i\ep-\ep_\gamma(\ep,\bk)\;}\\
&\qquad
\partial_\mu G_*(i\ep,\bk)^{-1}_{\alpha\beta}\;\partial_\mu G_*(i\ep,\bk)^{-1}_{\beta\gamma}
\;\partial_\mu G_*(i\ep,\bk)^{-1}_{\gamma\alpha}\;,
\label{0}
\eal
having defined    
\bealn
&\partial_\mu G_*(i\ep,\bk)^{-1}_{\alpha\beta} \equiv \bra{\alpha(\ep,\bk)}\partial_\mu G_*(i\ep,\bk)^{-1}
\ket{\beta(\ep,\bk)}\\
&\quad = i\,\delta_{\mu 0}\,\delta_{\alpha\beta} -\bra{\alpha(\ep,\bk)}\partial_\mu H_*(\ep,\bk)
\ket{\beta(\ep,\bk)}\\
&\quad \equiv i\,\delta_{\mu 0}\,\delta_{\alpha\beta} - F^\mu_{\alpha\beta}(\ep,\bk)\,.
\eal
The term in \eqn{0} with $\alpha=\beta=\gamma$ vanishes because of the antisymmetric tensor, so that 
we are left with the cases of either two states equal and different from the third, 
or of all states different, which we denote as $W^{(1)}(G)$ and  
$W^{(2)}(G)$, respectively. Specifically, 
\beal
&W^{(1)}(G) =-\fract{1}{8\pi^2}\,\int d\ep\,d\bk\,\ep_{\mu\nu\rho}
\\
&\qquad \sum_{\alpha\beta}{'}\,
\fract{1}{\;i\ep-\ep_\beta(\ep,\bk)\;}\;
\partial_\mu\,\Bigg(\fract{1}{\;i\ep-\ep_\alpha(\ep,\bk)\;}\Bigg)\\
&\qquad \qquad \qquad \qquad\qquad
 \bigg\{F^\nu_{\alpha\beta}(\ep,\bk)\,
F^\rho_{\beta\alpha}(\ep,\bk)\bigg\}\,,\\
&W^{(2)}(G) =-
\fract{1}{24\pi^2}\,\int d\ep\,d\bk\,\ep_{\mu\nu\rho}\\
&\quad \sum_{\alpha\beta\gamma}{'}\,\fract{1}{\;i\ep-\ep_\alpha(\ep,\bk)\;}
\fract{1}{\;i\ep-\ep_\beta(\ep,\bk)\;}\fract{1}{\;i\ep-\ep_\gamma(\ep,\bk)\;}\\
&\qquad\qquad \qquad \qquad
\bigg\{F^\mu_{\alpha\beta}(\ep,\bk)\;F^\nu_{\beta\gamma}(\ep,\bk)\;F^\rho_{\gamma\alpha}(\ep,\bk)\bigg\}\,,
\label{W1 and W2}
\eal
where $\Sigma{'}$ means the summation over different indices.\\ 
Let us begin by analysing $W^{(1)}(G)$ in Eq.~\eqn{W1 and W2}. 
We note that, for $\alpha\not=\beta$, 
\beal
&F^\nu_{\alpha\beta}(\ep,\bk) =\bra{\alpha(\ep,\bk)}\partial_\nu H_*(\ep,\bk)
\ket{\beta(\ep,\bk)} \\
&\quad = \Big(\ep_\alpha(\ep,\bk)-\ep_\beta(\ep,\bk)\Big)\,\bra{\partial_\nu\alpha(\ep,\bk)}
\beta(\ep,\bk)\rangle\\
&\quad =\Big(\ep_\beta(\ep,\bk)-\ep_\alpha(\ep,\bk)\Big)\,\bra{\alpha(\ep,\bk)}
\partial_\nu\beta(\ep,\bk)\rangle\,,
\label{F}
\eal
so that 
\beal 
&W^{(1)}(G) = -\fract{1}{8\pi^2}\,\int d\ep\,d\bk\,\ep_{\mu\nu\rho}
\\
&\quad \sum_{\alpha\beta}\,
\fract{\;\big(\ep_\alpha(\ep,\bk)-\ep_\beta(\ep,\bk)\big)^2\;}
{\;i\ep-\ep_\beta(\ep,\bk)\;}\;
\partial_\mu\,\bigg(\fract{1}{\;i\ep-\ep_\alpha(\ep,\bk)\;}\bigg)\\
&\qquad\quad 
\bra{\partial_\nu\alpha(\ep,\bk)}\beta(\ep,\bk)\rangle
\bra{\beta(\ep,\bk)}\partial_\rho \alpha(\ep,\bk)\rangle\\
&\quad 
= -\fract{1}{16\pi^2}\,\int d\ep\,d\bk\,\ep_{\mu\nu\rho}\,\sum_{\alpha\beta}\,
\partial_\mu\,S_{\alpha\beta}(\ep,\bk)\\
&\qquad\qquad  \bra{\partial_\nu\alpha(\ep,\bk)}\beta(\ep,\bk)\rangle
\bra{\beta(\ep,\bk)}\partial_\rho \alpha(\ep,\bk)\rangle\,,
\label{3}
\eal
where the constraint $\alpha\not=\beta$ is automatically fulfilled and one can readily demonstrate that 
\bealn
S_{\alpha\beta}(\ep,\bk)&= 2\,\ln\fract{\;i\ep-\ep_\alpha(\ep,\bk)\;}{\;i\ep-\ep_\beta(\ep,\bk)\;}
-\fract{\;i\ep-\ep_\alpha(\ep,\bk)\;}{\;i\ep-\ep_\beta(\ep,\bk)\;}\\
&\qquad +\fract{\;i\ep-\ep_\beta(\ep,\bk)\;}{\;i\ep-\ep_\alpha(\ep,\bk)\;}\,,
\eal
which has a discontinuous imaginary part crossing $\ep=0$ if 
$\ep_\alpha(0,\bk)\,\ep_\beta(0,\bk)<0$. We can write
\bealn
S_{\alpha\beta}(\ep,\bk)&= K_{\alpha\beta}(\ep,\bk) \\
&\qquad +2\pi i\,\text{sign}(\ep)\,\Big[\theta\big(\ep_\alpha(\ep,\bk)\big)
-\theta\big(\ep_\beta(\ep,\bk)\big)\Big]\,,
\eal
where $K_{\alpha\beta}(\ep,\bk)$ is now continuous at $\ep=0$, so that, since $S_{\alpha\beta}(\ep,\bk)$ 
is antisymmetric, and 
$\ep_\alpha(\ep,\bk)\not=0$, $\forall\,\alpha,\ep,\bk$, then  
\beal 
&W^{(1)}(G) = \fract{i}{2\pi}\,\int d\bk\,\ep_{ij}\,\sum_{\alpha}\,
\theta\big(-\ep_\alpha(0,\bk)\big)\\
&\qquad\qquad\qquad\qquad\qquad \bra{\partial_i\alpha(0,\bk)}\partial_j \alpha(0,\bk)\rangle\\
&\qquad\qquad  -\fract{1}{16\pi^2}\,\int d\ep\,d\bk\,\ep_{\mu\nu\rho}\,\sum_{\alpha\beta}\,
\partial_\mu\,K_{\alpha\beta}(\ep,\bk)\\
&\qquad\qquad\; \bra{\partial_\nu\alpha(\ep,\bk)}\beta(\ep,\bk)\rangle
\bra{\beta(\ep,\bk)}\partial_\rho \alpha(\ep,\bk)\rangle\,,
\label{4}
\eal
where $i,j=1,2$. The second term, which we denote as $I$, is only contributed by 
$\Ima\,K_{\alpha\beta}(\ep,\bk)$, which is odd in $\ep$, vanishes at $\ep\to \pm\infty$ and, by definition, is continuous at $\ep=0$. That allows partial integration, which, through Eq.~\eqn{F}, leads to
\beal
I &= -\fract{1}{8\pi^2}\,\int d\ep\,d\bk\,\ep_{\mu\nu\rho}\,\sum_{\alpha\beta}\,
K_{\alpha\beta}(\ep,\bk)\\
&\quad 
\bra{\partial_\mu\alpha(\ep,\bk)}\partial_\nu\beta(\ep,\bk)\rangle
\bra{\beta(\ep,\bk)}\partial_\rho \alpha(\ep,\bk)\rangle\\
&= -\fract{1}{8\pi^2}\int d\ep\,d\bk\,\ep_{\mu\nu\rho}\,
\sum_{\alpha\beta\gamma}\,
\fract{K_{\alpha\beta}(\ep,\bk)}{\;\ep_\alpha(\ep,\bk)-\ep_\gamma(\ep,\bk)\;}
\\
&\qquad \fract{1}{\;
\ep_\beta(\ep,\bk)-\ep_\gamma(\ep,\bk)\;}\,
\fract{1}{\;\ep_\alpha(\ep,\bk)-\ep_\beta(\ep,\bk)\;}\\
&\qquad\qquad \bigg\{F^\mu_{\alpha\gamma}(\ep,\bk)
\,F^\nu_{\gamma\beta}(\ep,\bk)\,F^\rho_{\beta\alpha}(\ep,\bk)\bigg\}\,.
\label{I-1}
\eal
Since $I$ is real, we can take the complex conjugate, send $\ep\to-\ep$ and then either exchange $\beta$ and $\gamma$ 
as well as $\mu$ and $\rho$, thus getting 
\beal
I &= -\fract{1}{8\pi^2}\int d\ep\,d\bk\,\ep_{\mu\nu\rho}\,
\sum_{\alpha\beta\gamma}\,
\fract{-K_{\alpha\gamma}(\ep,\bk)}{\;\ep_\alpha(\ep,\bk)-\ep_\gamma(\ep,\bk)\;}
\\
&\qquad \fract{1}{\;
\ep_\beta(\ep,\bk)-\ep_\gamma(\ep,\bk)\;}\,
\fract{1}{\;\ep_\alpha(\ep,\bk)-\ep_\beta(\ep,\bk)\;}\\
&\qquad\qquad \bigg\{F^\mu_{\alpha\gamma}(\ep,\bk)
\,F^\nu_{\gamma\beta}(\ep,\bk)\,F^\rho_{\beta\alpha}(\ep,\bk)\bigg\}\,,
\label{I-2}
\eal
or, instead, exchange $\alpha$ and $\gamma$ 
as well as $\nu$ and $\rho$, in that way obtaining  
\beal
I &= -\fract{1}{8\pi^2}\int d\ep\,d\bk\,\ep_{\mu\nu\rho}\,
\sum_{\alpha\beta\gamma}\,
\fract{-K_{\gamma\beta}(\ep,\bk)}{\;\ep_\alpha(\ep,\bk)-\ep_\gamma(\ep,\bk)\;}
\\
&\qquad \fract{1}{\;
\ep_\beta(\ep,\bk)-\ep_\gamma(\ep,\bk)\;}\,
\fract{1}{\;\ep_\alpha(\ep,\bk)-\ep_\beta(\ep,\bk)\;}\\
&\qquad\qquad \bigg\{F^\mu_{\alpha\gamma}(\ep,\bk)
\,F^\nu_{\gamma\beta}(\ep,\bk)\,F^\rho_{\beta\alpha}(\ep,\bk)\bigg\}\,.
\label{I-3}
\eal
Therefore, recalling that $K_{\alpha\beta}(\ep,\bk)=-K_{\beta\alpha}(\ep,\bk)$ is antisymmetric, we can rewrite $I$ as one third of the sum of \eqn{I-1}, \eqn{I-2}
and \eqn{I-3}, thus   
\bw
\bealn 
I &= \fract{1}{24\pi^2}\,\int d\ep\,d\bk\,\ep_{\mu\nu\rho}\,\sum_{\alpha\beta\gamma}\,
\fract{K_{\alpha\beta}(\ep,\bk)+K_{\beta\gamma}(\ep,\bk)+K_{\gamma\alpha}(\ep,\bk)}{\;
\big(\ep_\alpha(\ep,\bk)-\ep_\beta(\ep,\bk)\big)
\big(\ep_\beta(\ep,\bk)-\ep_\gamma(\ep,\bk)\big)
\big(\ep_\gamma(\ep,\bk)-\ep_\alpha(\ep,\bk)\big)
\;}\\
&\qquad\qquad\qquad\qquad\qquad\qquad \bigg\{ F^\mu_{\alpha\gamma}(\ep,\bk)
\,F^\nu_{\gamma\beta}(\ep,\bk)\,F^\rho_{\beta\alpha}(\ep,\bk)\bigg\}\\
&= \fract{1}{24\pi^2}\,\int d\ep\,d\bk\,\ep_{\mu\nu\rho}\,\sum_{\alpha\beta\gamma}{'}\,\fract{1}{\;i\ep-\ep_\alpha(\ep,\bk)\;}
\fract{1}{\;i\ep-\ep_\beta(\ep,\bk)\;}\fract{1}{\;i\ep-\ep_\gamma(\ep,\bk)\;}\,\bigg\{F^\mu_{\alpha\beta}(\ep,\bk)\;F^\nu_{\beta\gamma}(\ep,\bk)\;F^\rho_{\gamma\alpha}(\ep,\bk)\bigg\}\\
&= -W^{(2)}(G)\,,
\eal
\ew
where the equivalence between the first and the second equations can be readily worked out.\\
In conclusion, we have proved that the winding number \eqn{W} can be 
written, not unexpectedly, as 
\bealn 
W(G) &= -\fract{1}{16\pi^2}\,\int d\ep\,d\bk\,\ep_{\mu\nu\rho}\,
\partial_\mu\,\Bigg\{\sum_{\alpha\beta}
\,S_{\alpha\beta}(\ep,\bk)\\
&\qquad  \bra{\partial_\nu\alpha(\ep,\bk)}\beta(\ep,\bk)\rangle
\bra{\beta(\ep,\bk)}\partial_\rho \alpha(\ep,\bk)\rangle\Bigg\}\,,
\eal
namely, as the integral of a full derivative of a function 
that has a discontinuity at $\ep=0$, for which reason the integral does not vanish and yields 
\beal 
W(G) &= \fract{i}{2\pi}\,\int d\bk\,\ep_{ij}\,\sum_{\alpha}\,
\theta\big(-\ep_\alpha(0,\bk)\big)\\
&\qquad\qquad\qquad\qquad  \bra{\partial_i\alpha(0,\bk)}\partial_j \alpha(0,\bk)\rangle\,,
\label{final-W}
\eal
i.e., the TKNN expression~\cite{TKNN} for \textit{quasiparticles} 
described by the non-interacting Hamiltonian $H_*(0,\bk)$. Since $H_*(\ep,\bk)$ 
includes by definition both poles and zeros of the retarded Green's function, 
we conclude that both of them contribute on equal footing to the topological invariant $W(G)$.\\
We emphasise that only through the representation \eqn{Green's} of the Green's 
function $G(i\ep,\bk)$ we have been able to straightforwardly derive the simple 
expression of $W(G)$ in Eq.~\eqn{final-W}. Since that same representation is also the key to the proof that Landau's quasiparticles exist at Luttinger 
as well as Fermi surfaces~\cite{mio-2,mio-Mott}, we suspect it is not just a mathematical trick but hints at a deeper physical meaning. Indeed, we are convinced that the bands of $H_*(0,\bk)$ lying inside the single-particle gap 
of a Mott insulator describe \textit{fractionalised} quasiparticles not carrying all electron's quantum numbers, for instance 
neutral but spinful, even when they do not cross the chemical potential. Moreover, since those fractionalised quasiparticles cannot have any weight in the physical 
single-particle excitations, it is reasonable to expect that they are associated 
to the existence of ingap bands of zeros of the physical retarded Green's function.\\

\noindent
Finally, we remark that all above results has been obtained in two dimensions (2D). 
However, the expression of topological invariants in 3D insulators~\cite{Zhang-PRL2010,Vanderbilt-PRB2010,Moore-ARCMP2011,Zhang-PRX2012,Gabi-PRB2013,resta2020geometry} 
also involve winding numbers that generalise \eqn{W} in higher dimensions~\cite{Zhang-PRX2012}, or describe other invariants like polarisation~\cite{Gabi-PRB2013}. Therefore, it remains true that the quasiparticle residue \eqn{Z} disappears from the expression of the topological invariant, which is therefore 
only determined by the quasiparticle Green's function $G_*(i\ep,\bk)$, see 
Eq.~\eqn{Green's}, exactly like in 2D. For instance, we can readily 
show, simply following \cite{Zhang-PRB2012}, that the parity of the eigenvalues of 
$H_*(0,\bk)$ at time-reversal invariant momenta determine 
the topological 
invariant~\cite{Zhang-PRL2010} of interacting $Z_2$ topological insulators with inversion symmetry. 

\section{A toy example}
\label{A toy example}
We now analyse a toy model inspired by Ref.~\cite{Giorgio}. Specifically, we consider an interacting BHZ model~\cite{BHZ}, whose inverse Green's function for a fixed spin reads  
\beal
\hat{G}(i\ep,\bk)^{-1} &= i\ep -\ep(\bk)\,\hat{\tau}_3 - \lambda\,\sin k_1\,\hat{\tau}_1\\
&\qquad +\lambda\,\sin k_2\,\hat{\tau}_2 -\hat{\Sigma}(i\ep,\bk)\\
&= i\ep -\hat{H}(\bk) -\hat{\Sigma}(i\ep,\bk)\,,
\label{H-BHZ}
\eal
where $\ep(\bk) = M -\cos k_1 -\cos k_2$, a hat is introduced to distinguish matrices from scalars, and $\hat{\tau}_a$, $a=1,2,3$, are the Pauli matrices in the two-orbital subspace.  Without interaction, the model describes 
a topological insulator if $\ep(\bd{\Gamma})\,\ep(\bf{M}) < 0$, with $\bf{M}=(\pi,\pi)$, which occurs if $0<|M| <2 $.\\
We assume that~\cite{Giorgio} 
\beal
\hat{\Sigma}(i\ep,\bk) &= \fract{\Delta^2}{\;i\ep + \hat{H}'(\bk)\;}\;,
\label{Sigma-BHZ}
\eal
with $\Delta>0$ and where $\hat{H}'(\bk)$ has the same form as $\hat{H}(\bk)$ in \eqn{H-BHZ} but with renormalised parameters, thus $\ep(\bk)\to \ep'(\bk) = M' -t'\big(\cos k_1 +\cos k_2\big)$ and $\lambda\to\lambda'$. It follows that $\hat{H}'(\bk)$   
is topological if $0<|M'| <2 $, which we take for granted. \\
We also assume that the model is deep inside the Mott insulating regime, which implies 
that $\Delta$ is in magnitude much larger than all the other parameters in 
\eqn{H-BHZ}. In this case, the poles of the retarded Green's function, $\hat{G}(\ep+i0^+,\bk)$, describes two lower and two upper Hubbard bands, with dispersion, respectively, $\ep_\text{LHB}(\bk)\simeq -\Delta + \delta\ep_{1(2)}(\bk)\ll 0$ 
and $\ep_\text{UHB}(\bk)\simeq +\Delta + \delta\ep_{1(2)}(\bk)\gg 0$, where 
$\delta\ep_{1(2)}(\bk)$ are the eigenvalues of $\hat{H}(\bk)-\hat{H}'(\bk)$. 
The occupied lower Hubbard bands have opposite Chern numbers so that, from the point of view of the Green's function poles, the system is a trivial Mott insulator, as noted in \cite{Giorgio}.\\
However, besides those poles, the retarded Green's function also has valence and  conduction bands of zeros with dispersion the eigenvalues of $-\hat{H}'(\bk)$ in Eq.~\eqref{Sigma-BHZ}, which are therefore topological~\cite{Giorgio}.  \\

\noindent
The obvious question is whether the non-trivial topology of the Green's function zeros has any physical significance. For that, we follows the analysis of the previous section.    
Upon defining $E'(\bk)^2= \ep'(\bk)^2 + \lambda'{^2}\,\big(\sin^2 k_1+ \sin^2 k_2\big)$ one readily finds that the quasiparticle residue \eqn{Z} reads in this case
\beal
\hat{Z}(\ep,\bk) &= \fract{\;\ep^2 + E'(\bk)^2\;}{\;\ep^2 + E'(\bk)^2 + \Delta^2\;} = Z(\ep,\bk)\,\hat{I}\;,
\eal
and is proportional to the identity matrix $\hat{I}$, and thus 
\bealn
\hat{H}_*(\ep,\bk) &= Z(\ep,\bk)\,
\Bigg(\hat{H}(\bk) + \fract{\Delta^2}{\;\ep^2 + E'(\bk)^2\;}\;\hat{H}'(\bk)\Bigg)
\,,
\eal
which has exactly the same form as $\hat{H}$ with frequency dependent parameters 
$\ep_*(\ep,\bk)$ and $\lambda_*(\ep,\bk)$ that can be easily determined. 
At $\ep=0$ and for large $\Delta$,
\bealn
\hat{H}_*(0,\bk) &= Z(0,\bk)\,
\Bigg(\hat{H}(\bk) + \fract{\Delta^2}{\;E'(\bk)^2\;}\;\hat{H}'(\bk)\Bigg)
\simeq \hat{H}'(\bk)\,,
\eal
explicitly showing that only the Green's function zeros contribute to the topological invariant \eqn{final-W} in this toy example, and with opposite sign respect to the topology of the valence band of zeroes that are described by $-\hat{H}'$. We emphasise that the exact correspondence between 
the ingap quasiparticle bands, eigenvalues of the Hamiltonian $\hat H_*(0,\bk)$, and the inverted bands of zeros of the retarded physical Green's function holds only in the limit of 
infinite Mott-Hubbard gap.

\section{Concluding remarks}
The winding number $W(G)$ of the physical electron Green's function $G(i\ep,\bk)\in \text{GL}(n,\mathbb{C})$ can be written as the winding 
number $W(G_*)$ of a quasiparticle Green's function 
$G_*(i\ep,\bk) = 1/\big(i\ep-H_*(\ep,\bk)\big)$, see Eq.~\eqn{Green's}, whose poles on the real frequency 
axis are associated to both poles and zero of $G(\ep,\bk)$. We have shown explicitly that $W(G_*)$ reduces to the famous TKNN formula for free electrons, here the quasiparticles, described by the Hamiltonian $H_*(0,\bk)$. \\

\noindent
This result implies that, against all expectations, 
the zeros of the real frequency Green's function do have a topological relevance, which is consistent with earlier studies~\cite{Volovik-1,Gurarie-zeros-PRB2011,Essing&Gurarie-PRB2011,mio-2,Jan,mio-Mott,Giorgio}, and nonetheless striking. Indeed, one would na\"{\i}vely argue that the position of the ingap zeros could be easily changed from the positive to the negative side of the real frequency axis, or vice versa, by slightly modifying 
the Hamiltonian parameters, e.g., moving the chemical potential inside the insulating gap. 
However, if one accepts our viewpoint that ingap bands of zeros, or, more correctly, ingap 
bands of the quasiparticle Hamiltonian $H_*(0,\bk)$, may describe genuine excitations that do not carry 
all electron's quantum numbers, then their response to a shift in chemical potential is expected 
to differ substantially from that of non-interacting bands. \\
An enlightening example is in our opinion offered by a  
Hubbard atom with Hamiltonian $U(n-1)^2/2$, the simplest realisation of a Mott insulator. Its Green's function on the real frequency axis, 
\bealn
G(\ep) = \fract{1}{2}\,\left(
\fract{1}{\;\ep+U/2\;} + \fract{1}{\;\ep-U/2\;}\right)\,,
\eal
has poles at $\ep = \pm U/2$ and a zero at $\ep=0$. Through equations \eqn{Green's} and \eqn{Z} 
one finds that, for imaginary frequencies,  
\bealn 
Z(\ep) &= \fract{\ep^2}{\;\ep^2+U^2/4\;}\;,& G_*(i\ep) &= \fract{1}{\;i\ep\;}\;.
\eal
It is tempting to associate the zero-frequency pole of the quasiparticle $G_*(i\ep)$ and 
the vanishing quasiparticle residue $Z(0)=0$ 
to the free spin-1/2 of the isolated atom. In contrast, a non-interacting atom, $U=0$, has
\bealn
G(i\ep) &= G_*(i\ep) = \fract{1}{\;i\ep\;}\;,& Z(\ep) &= 1\,,
\eal
consistently with the fact that the zero frequency excitations are physical single-particle ones, $Z(0)=1$. 
We now imagine to couple the atom to a metallic reservoir 
with which it exchanges electrons, as one would do in statistical mechanics to fix the chemical potential, 
and assume that the atomic level is at energy $\ep_d$ with respect to the chemical potential of the reservoir. 
The atom plus the reservoir thus describe a conventional Anderson impurity model. 
In the case of the Hubbard atom with $\ep_d\ll U$, one expects that the zero-frequency pole of the 
quasiparticle Green's function is immediately promoted to a Kondo resonance pinned at the bath chemical potential, thus 
\bealn
G_*(i\ep) &=\fract{1}{\;i\ep\;} \to  
 \fract{1}{\;i\ep + i\,T_K\,\text{sign}(\ep)\;}\;,& 
 Z(\ep) &\to \fract{T_K}{\Gamma}\;,
 \eal
where $\Gamma$ is the \textit{bare} hybridisation width and $T_K$ the Kondo temperature. In other words, the coupling to the bath allows revealing the 
hidden physical meaning of the zero, i.e., its being a free spin prompt to Kondo screening. 
On the contrary, in the case of the non-interacting atom 
\bealn
G_*(i\ep) =\fract{1}{\;i\ep\;} \to  
 \fract{1}{\;i\ep - \ep_d + i\,\Gamma\,\text{sign}(\ep)\;}\;,
 \eal
which describes a resonant level centred at $\ep_d$. It is remarkable that, while the non-interacting 
atom simply inherits the chemical potential of the bath, the Hubbard atom does 
not; the Kondo resonance is always pinned at the chemical potential even though the atomic level is offset by $\ep_d$.\\
This very simple example not only supports our 
interpretation that ingap quasiparticle bands in Mott insulators may describe 
fractionalised excitations, but also 
suggests that these bands respond very differently from conventional ones 
to a change in chemical potential induced by the contact with a charge reservoir, 
in contrast to a recent claim~\cite{Phillips-boh}.  \\

\noindent
Our analysis also extends the notion of topological transitions and adiabatic transformations for strongly interacting electrons. Indeed, it was already known that some topological invariants are contributed by zero-frequency roots of the Green function~\cite{Volovik-1}. That, however, seems at odds with the expected behaviour of topological invariants under adiabatic transformations, since neither a closure of the charge gap nor any symmetry breaking occurs when 
Green's function zeros cross the chemical potential.
In light of our results, this phenomenon acquires a straightforward physical explanation: 
Green's function zeros crossing the chemical potential form a Luttinger surface that hosts gapless excitations~\cite{mio-2,mio-Mott} despite the finite charge gap, thus providing  the non-adiabaticity required to change topology.\\ 

\noindent
Finally, our results raise several questions worth being addressed in the future. 
The correspondence between topological bands and edge modes 
of Green's function zeros~\cite{Gurarie-zeros-PRB2011,Essing&Gurarie-PRB2011,Giorgio} suggests that, similarly to the conventional case of edge poles, the edge zeros are ultimately responsible of the quantised Hall conductance 
\eqn{final-W}, although the surface is charge insulating. 
In the model quantum spin-Hall insulator of Sec.~\ref{A toy example}, 
that puzzling prediction can be explained by noticing that, according to 
Ref.~\cite{mio-Mott}, the edge quasiparticles at the Luttinger surface, actually a point, can carry a spin current, thus a quantised spin-Hall conductance. 
However, that simple explanation would not work for a hypothetical Chern Mott insulator with edge zeros crossing the chemical potential. Therefore, even though the results of Sec.~\ref{TKNN formula for interacting insulators} prove that 
bulk bands of Green's function zeros contributes to the topological invariant \eqn{W}, the actual role of edge zeros remains unclear. 
\\
A further question regards fractional Chern insulators. 
Indeed, if the winding number \eqn{W} does correspond to the Hall conductance, 
which may not always be the case, see Appendix 
\ref{Topological invariant in two dimensions}, one may wonder how 
it may ever be fractional since the TKNN formula should yield an integer value. In view of the similar issue that arises in the fractional quantum Hall effect~\cite{Thouless-PRB1985}, we believe that the ground state degeneracy, also expected in a fractional Chern insulator~\cite{Bernevig-PRX2011}, is the key ingredient. Specifically, we suspect that $H_*(0,\bk)$ calculated over each ground state has valence bands with ill-defined Chern number, because, e.g., they touch the zone boundaries with finite slope. However, assuming, for simplicity, that the ground state is threefold degenerate, 
the Hall conductance is better defined as~\cite{Thouless-PRB1985}
\bealn
\sigma_\text{H} & = \fract{1}{3}\,\Big(W(G_1)+W(G_2)+W(G_3)\Big) \\
&= \fract{1}{3}\,W\big(G_1\,G_2\,G_3\big)\,,
\eal
where $G_n$ is the Green's function calculated over the ground state $\ket{n}$, $n=1,2,3$. 
The sum of the three winding numbers corresponds to the TKNN formula applied to the quasiparticles 
valence bands of all three ground states. We speculate, see, e.g., Ref.~\cite{Zubkov-boh-2021}, that all these bands as a whole correspond to a well-behaved single band once unfolded into a threefold larger Brillouin zone, whose Chern number is therefore 
an integer $\ell$, thus $\sigma_\text{H}=\ell/3$, a fractional value. Incidentally, it is suggestive that the sum of the winding numbers is just the winding number 
of $G=G_1\,G_2\,G_3$, which is still a complex invertible matrix. The above is just 
a speculation that we believe worth investigating. \\

\noindent
We conclude by emphasising that the winding number \eqn{W} in two dimensions, although being a topological invariant, not necessarily coincides with the quantised Hall conductance when perturbation theory breaks down, see Appendix 
\ref{Topological invariant in two dimensions}. Similarly, we cannot exclude 
that the extensions of the Green's function winding number in three 
dimensions~\cite{Zhang-PRL2010,Zhang-PRX2012,Zhang-PRB2012,Gabi-PRB2013}
might be unrelated to the physical observables they are supposed 
to reproduce when there is no adiabatic connection between the interacting system 
and the non-interacting one. That leaves open the question about the actual physical meaning of those winding numbers~\cite{Slage-PRB2015,He-PRB2016-II}.    

\begin{acknowledgments}
We are very grateful to Adriano Amaricci, Elio K\"onig
Michel Ferrero, Giorgio Sangiovanni 
and, especially, Roberto Percacci for helpful discussions and comments. 
\end{acknowledgments}

\appendix

\section{Properties of thermal Green's functions}

\label{Properties of thermal Green's functions}
Hereafter, we consider a generic basis of Bloch wavefunctions 
$\phi_{a\bk}(\bk)$, where $a$ include also the spin label, with associated 
creation, $c^\dagger_{a\bk}$, and annihilation, $c^\dagga_{a\bk}$, operators. We assume that lattice translational symmetry is not broken. The Green's function matrix in imaginary time $\tau\in[-\beta,\beta]$, where 
$\beta=1/T$, is defined through its components, diagonal in momentum, 
\beal
G_{ab}(\tau,\bk) &= -\big\langle\,T_\tau\Big(c^\dagga_{a\bk}(\tau)\,c^\dagger_{b\bk}\Big)\,\big\rangle\,,
\label{G(tau)}
\eal
where $T_\tau$ is the time-ordering operator and $c^\dagga_{a\bk}(\tau) = 
\esp{H\tau}\;c^\dagga_{a\bk}\;\esp{-H\tau}$ the imaginary time evolution of the operator. 
The Fourier transform of \eqn{G(tau)} in Matsubara frequencies $i\ep_\ell=(2\ell+1)\pi T$ 
is readily found to be
\bealn
G_{ab}(i\ep_\ell,\bk) &= \fract{1}{\Omega}\,
\sum_{n,m}\,\bigg(\esp{-\beta E_n}+\esp{-\beta E_m}\bigg)\\
&\qquad\qquad\quad \fract{\;\bra{n}c^\dagga_{a\bk}\ket{m}\bra{m}c^\dagger_{b\bk}\ket{n}\;}
{\;i\ep_\ell -\big(E_m-E_n\big)\;}\;,
\eal
where $\Omega=\Tr\left(\esp{-\beta H}\right)$ is the partition function, $\ket{\!\!n}$ and $\ket{\!\!m}$ many-body eigenstates with eigenvalues $E_n$ and $E_m$, respectively. We note that 
$G_{ab}(i\ep_\ell,\bk)^* = G_{ba}(-i\ep_\ell,\bk)$, and thus 
\beal
G(i\ep_\ell,\bk)^\dagger = G(-i\ep_\ell,\bk)\,.
\label{G-property}
\eal
This property must evidently hold true also for the inverse matrices. Therefore, using Dyson's equation
\bealn
G(i\ep_\ell,\bk)^{-1} &= i\ep_\ell - H_0(\bk) - \Sigma(i\ep_\ell,\bk)\,,
\eal 
where $H_0(\bk)$ is the non-interacting Hamiltonian that includes the chemical potential term $-\mu\,N$,  and $\Sigma(i\ep_\ell,\bk)$ the self energy, both being matrices in the same basis of Bloch wavefunctions, one concludes that also 
\beal
\Sigma(i\ep_\ell,\bk)^\dagger=\Sigma(-i\ep_\ell,\bk)\,.
\label{Sigma-property}
\eal  
We can write $G(i\ep_\ell,\bk) = G_1(i\ep_\ell,\bk)+i\,G_2(i\ep_\ell,\bk)$ where 
\bealn
G_1(i\ep_\ell,\bk) &= \fract{\;G(i\ep_\ell,\bk)+G(i\ep_\ell,\bk)^\dagger\;}{2}\\
&=\fract{\;G(i\ep_\ell,\bk)+G(-i\ep_\ell,\bk)\;}{2}\;,\\
G_2(i\ep_\ell,\bk) &= \fract{\;G(i\ep_\ell,\bk)-G(i\ep_\ell,\bk)^\dagger\;}{2i}\\
&=\fract{\;G(i\ep_\ell,\bk)-G(-i\ep_\ell,\bk)\;}{2i}\;,
\eal
are both hermitian, $G_1$ even and $G_2$ odd in $\ep_\ell$. Specifically, the matrix elements 
of $G_2$ read
\bealn
G_{2\,ab}(i\ep_\ell,\bk) &= -\fract{\;\ep_\ell\;}{\Omega}\,
\sum_{n,m}\,\bigg(\esp{-\beta E_n}+\esp{-\beta E_m}\bigg)\\
&\qquad\qquad\qquad\quad 
\fract{\;\bra{n}c^\dagga_{a\bk}\ket{m}\bra{m}c^\dagger_{b\bk}\ket{n}\;}
{\;\ep_\ell^2 +\big(E_m-E_n\big)^2\;}\;.
\eal
Since $G_2$ is hermitian, it can be diagonalised by a unitary transformation 
$c^\dagga_{a\bk}\to c^\dagga_{\alpha\bk}$ that yields the eigenvalues 
\bealn
G_{2\,\alpha}(i\ep_\ell,\bk) &= -\fract{\;\ep_\ell\;}{\Omega}\,
\sum_{n,m}\,\bigg(\esp{-\beta E_n}+\esp{-\beta E_m}\bigg)\\
&\qquad\qquad\qquad\quad
\fract{\;\big|\bra{n}c^\dagga_{\alpha\bk}\ket{m}\big|^2\;}
{\;\ep_\ell^2 +\big(E_m-E_n\big)^2\;}\;,
\eal
which are negative for $\ep_\ell >0$ and positive otherwise. It follows that, once we write 
$\Sigma(i\ep_\ell,\bk) = \Sigma_1(i\ep_\ell,\bk)+ i\,\Sigma_2(i\ep_\ell,\bk)$, where 
\bealn
\Sigma_1(i\ep_\ell,\bk) &= \fract{\;\Sigma(i\ep_\ell,\bk)+\Sigma(i\ep_\ell,\bk)^\dagger\;}{2}\\
&=\fract{\;\Sigma(i\ep_\ell,\bk)+\Sigma(-i\ep_\ell,\bk)\;}{2}\;,\\
\Sigma_2(i\ep_\ell,\bk) &= \fract{\;\Sigma(i\ep_\ell,\bk)-\Sigma(i\ep_\ell,\bk)^\dagger\;}{2i}\\
&=\fract{\;\Sigma(i\ep_\ell,\bk)-\Sigma(-i\ep_\ell,\bk)\;}{2i}\;,
\eal
are also both hermitian, the former even in $\ep_\ell$ and the latter odd, the 
eigenvalues of $\Sigma_2(i\ep_\ell,\bk)$ are negative for $\ep_\ell >0$ and positive otherwise, just like those of $G_2(i\ep_\ell,\bk)$. \\

\noindent
The quasiparticle residue matrix is defined by \eqn{Z}, i.e., 
\bealn
Z(\ep_\ell,\bk) &= Z(-\ep_\ell,\bk) =\Bigg(1 - \fract{\;\Sigma_2(i\ep_\ell,\bk)\;}{\ep_\ell}\Bigg)^{-1}\,,
\eal
and is therefore a semi-positive definite matrix that becomes strictly positive definite in insulators 
without Luttinger surfaces, where $\Sigma_2(0,\bk)=0$. Similarly, the quasiparticle Hamiltonian \eqn{H*}, i.e., 
\bealn
&H_*(\ep_\ell,\bk) = H_*(-\ep_\ell,\bk)\\
&\quad =\sqrt{Z(\ep_\ell,\bk)\;}\;\Big(H_0(\bk) + \Sigma_1(i\ep_\ell,\bk)\Big)\;
\sqrt{Z(\ep_\ell,\bk)\;}\;,
\eal
is hermitian and even in $\ep_\ell$. Therefore, if $\ket{\alpha(\ep_\ell,\bk)}$ is eigenstate  
of $H_*(\ep_\ell,\bk)$ with eigenvalue $\ep_\alpha(\ep_\ell,\bk)$, we can always define 
$\ket{\alpha(-\ep_\ell,\bk)}=\ket{\alpha(\ep_\ell,\bk)}$ the eigenstate  
of $H_*(-\ep_\ell,\bk)$ with the same eigenvalue $\ep_\alpha(\ep_\ell,\bk)$. 
We end emphasising that the representation \eqn{Green's} of the Green's function 
in terms of the quasiparticle one,
\bealn
G_*(i\ep,\bk) = \fract{1}{\;i\ep-H_*(\ep,\bk)\;}\;,
\eal
with hermitian $H_*(\ep,\bk)$ is a rigorous result that remains valid also 
when $H_*(0,\bk)$ has zero eigenvalues, i.e., when Fermi and/or Luttinger surfaces are present. 

\section{Topological invariant in two dimensions} \label{Topological invariant in two dimensions}

The Hall conductance in non-interacting insulators is quantized in units $e^2/2\pi\hbar$, where the integer quantum is a topological invariant known as the first Chern number. This is calculated upon the occupied bands and is robust under any smooth deformation of the Hamiltonian. On the other hand, in the case of interacting electrons, band theory of independent electrons does not hold and yet a quantized topological invariant, which reduces to the Chern number if the interaction is switched off, can be still defined. However, while the expression of this invariant, the winding number \eqn{W}, could be well anticipated by algebraic topology arguments, we believe that the most accepted derivation, see, e.g., Refs.~\cite{Ishikawa-1, Ishikawa-2}, is not formally correct. 
Therefore, we here rederive the Hall conductance by standard quantum many-body theory~\cite{libro}. \\  
Hereafter, we consider a periodic model of interacting electrons and use units in which 
$\hbar=1$. 
The off-diagonal component $\sigma_{ij}$, $i\not=j$, 
of the conductivity tensor is related to the current-current response function 
$\chi_{ij}(\omega)$ through
\bealn
\sigma_{ij} &= -\lim_{\omega\to 0}\,\fract{e^2}{i\omega}\;\chi_{ij}(\omega)\,,
\eal
where, in the basis of the Hamiltonian eigenstates $\ket{n}$, with eigenvalues $E_n$,
\bealn
\chi_{ij}(\omega) &=\chi_{ji}(-\omega)=\fract{1}{Z}\,\sum_{nm}\,\fract{\;\esp{-\beta E_n}
-\esp{-\beta E_m}\;}{\omega-E_m+E_n}\\
&\qquad \int d\bx\,d\by\,
\bra{n}J_i(\bx)\ket{m}\bra{m}J_j(\by)\ket{n}\,,
\eal 
with $J_i(\bx)$ the $i$-the component of the current density operator at position $\bx$.
\begin{figure}[ht]
\centerline{\includegraphics[width=0.5\textwidth]{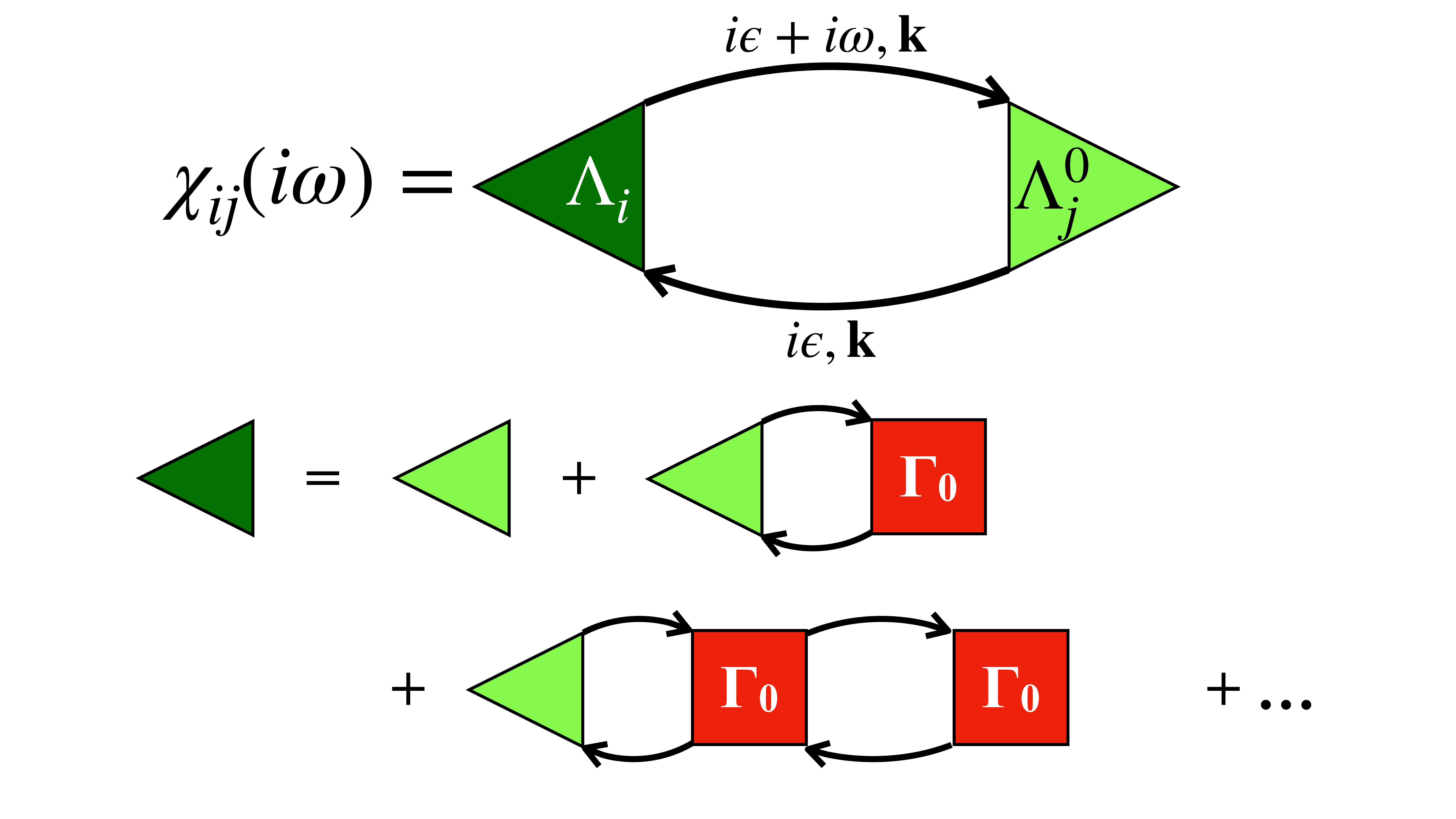}}
\caption{Top panel: current-current correlation function $\chi_{ij}(i\omega)$ in skeleton diagrams.
Black arrow lines are interacting single-particle Green's functions $G$, whose Matsubara frequency and momentum are explicitly shown; green triangle represents the non-interacting current vertex $\Lambda^0_j$, while dark green triangle the fully interacting one $\Lambda_i$. Bottom panel: Bethe-Salpeter equation satisfied by the interacting $\Lambda_i$ in terms of the interaction vertex $\Gamma_0$ irreducible in the 
particle-hole channel.}
\label{fig1}
\end{figure} 
Therefore, the antisymmetric component of the current-current tensor is,
for small $\omega$, 
\bealn
&\fract{\;\chi_{ij}(\omega) -\chi_{ji}(\omega)\;}{2} \simeq -\omega\;\fract{1}{Z}\,\sum_{nm}\fract{\;\esp{-\beta E_n}
-\esp{-\beta E_m}\;}{\big(E_m+E_n\big)^2}\\
&\qquad \int d\bx\,d\by\,
\bra{n}J_i(\bx)\ket{m}\bra{m}J_j(\by)\ket{n}\\
&= \omega\;\fract{\partial\chi_{ij}(\omega)}{\partial\omega}_{\big|\omega=0}\;,
\eal
which is imaginary, and thus, switching to Matsubara frequencies and for 
$i\not=j$, 
\bealn
\fract{\;\sigma_{ij} -\sigma_{ji}\;}{2} &= i\,e^2\,\fract{\partial\chi_{ij}(i\omega)}{\partial i\omega}_{\big|\omega=0}\;,
\eal
where now 
\bealn
\chi_{ij}(i\omega) &= -\int_0^\beta \!\! d\tau\,\esp{i\omega\tau}\!\!
\int \!d\bx\,d\by\,\langle\,T_\tau\Big(J_i(\bx,\tau)\,J_j(\by,0)\Big)\,\rangle\,,
\eal
is the correlation function in the Matsubara formalism, with $\tau$ the imaginary time. 
Fig.~\ref{fig1} shows the representation of $\chi_{ij}(i\omega)$ in skeleton diagrams. 
It is worth noticing that only one of the two current vertices is fully interacting, 
otherwise we would double count interaction effects as mistakenly done in Ref.~\cite{Ishikawa-2}. \\ 
We are interested in the derivative of $\chi_{ij}(i\omega)$ with respect to 
$i\omega$ calculated at $\omega=0$, which, by inspection of Fig.~\ref{fig1}, can be represented as 
in Fig.~\ref{fig2}, where, only because of the derivative, the current vertices are now both fully interacting. 
\begin{figure}[ht]
\centerline{\includegraphics[width=0.45\textwidth]{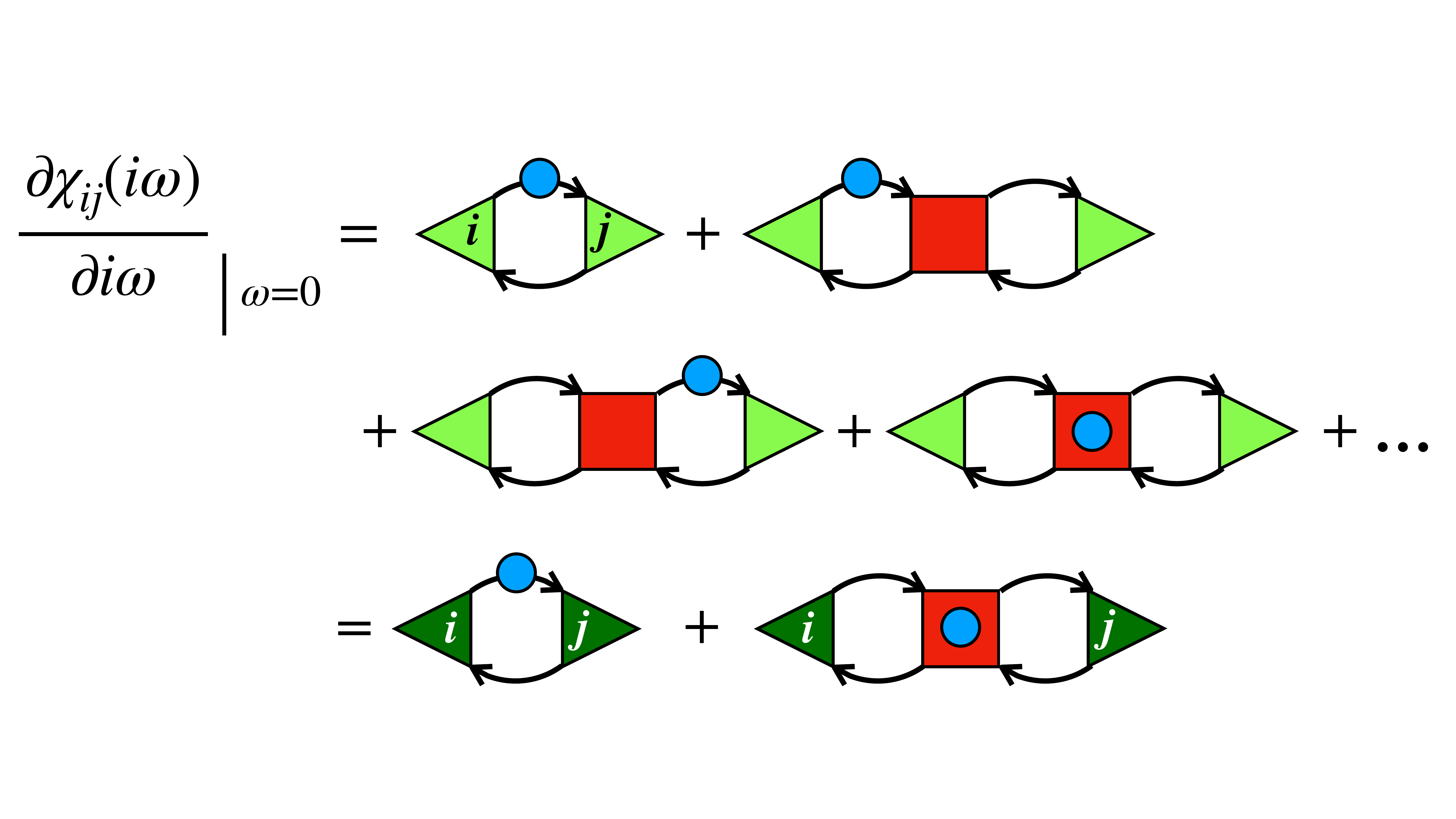}}
\vspace{-0.4cm}
\caption{Diagrammatic representation of $\partial\chi_{ij}(i\omega)/\partial i\omega$
up to first order in the skeleton expansion. The cyan circle represents $\partial/\partial i\omega$. Note that all diagrams must be evaluated at $\omega=0$ after taking the derivative.}
\label{fig2}
\end{figure} 
Through the Ward-Takahashi identity the fully interacting 
current vertex at $\omega=0$ is simply $\Lambda_i = -\partial G^{-1}/\partial k_i$, and, since $\partial G=-G\,\partial G^{-1}\,G$, we can write 
in 2D and for $T=0$ 
\bw
\beal
&\fract{\;\sigma_{12} -\sigma_{21}\;}{2} = \fract{e^2}{2\pi}\;\nu_\text{H}
\\
&\qquad =
\fract{e^2}{\left(2\pi\right)^3}\,
\int d\ep\,d\bk\,
\Tr\Big(G(i\ep,\bk)\;\partial_{k_1} G(i\ep,\bk)^{-1}\;
G(i\ep,\bk)\;\partial_\ep G(i\ep,\bk)^{-1}\;
G(i\ep,\bk)\;\partial_{k_2} G(i\ep,\bk)^{-1}\Big) +\fract{e^2}{2\pi}\; K_L\\
&\quad = \fract{e^2}{2\pi}\,\Bigg\{\fract{1}{24\pi^2}\,\int d\ep\,d\bk\,\ep_{\mu\nu\rho}\,
\Tr\Big(G(i\ep,\bk)\;\partial_\mu G(i\ep,\bk)^{-1}\;
G(i\ep,\bk)\;\partial_\nu G(i\ep,\bk)^{-1}\;
G(i\ep,\bk)\;\partial_\rho G(i\ep,\bk)^{-1}\Big) + K_L\Bigg\}\\
&\qquad = \fract{e^2}{2\pi}\,\Big[ W(G) + K_L\Big]\,,
\label{Hall}
\eal
\ew
where $\nu_H$ is the quantised Hall conductance, $\ep_{\mu\nu\rho}$, with indices running from 0 to 2, the antisymmetric tensor, $\partial_0=\partial_\ep$ and 
$\partial_{1(2)}=\partial_{k_{1(2)}}$, and $W(G)$, see \eqn{W}, is the winding number of the 
map $(\ep,\bk) \to G(i\ep,\bk)\in\text{GL}(n,\mathbb{C})$, assuming that 
the Green's function $G(i\ep,\bk)$ is an $n\times n$ invertible matrix, 
with $n$ the dimension of the single-particle wavefunction basis, which is true 
provided the system is insulating and has no Luttinger surface. \\
The additional term besides the winding number involves the derivative of the irreducible vertex $\Gamma_0$ and reads explicitly
\bw
\beal
K_L &= \fract{i\pi}{(2\pi)^6}\,\sum\,\int d\ep\, d\bk\,d\ep'\, d\bk'\;\ep_{ij}\;
\partial_i G_{ba}(i\ep,\bk)\;
F_{ab;a'b'}\big(i\ep,\bk;i\ep',\bk'\big)\;\partial_j G_{b'a'}(i\ep',\bk')\,,
\label{K_L}
\eal
\ew
where 
\bw
\beal
F_{ab;a'b'}\big(i\ep,\bk;i\ep',\bk'\big) &= \lim_{\omega\to 0}\,\fract{1}{2i\omega}\,
\bigg\{\Gamma_{0}\big(i\ep+i\omega\,\bk\,a,i\ep'\,\bk'\,a';
i\ep'+i\omega\,\bk'\,b',i\ep\,\bk\, b\big)\\
&\qquad\qquad\qquad\quad -\Gamma_{0}\big(i\ep\,\bk\,a,i\ep'+i\omega\,\bk'\,a';i\ep'\,\bk'\,b',
i\ep+i\omega\,\bk\,b\big)\Big\} = -F_{a'b';ab}\big(i\ep',\bk';i\ep,\bk\big)\,,
\label{F-Gamma}
\eal
\ew
with 
\bealn
&\Gamma_{0}\big(i\ep+i\omega\,\bk\,a,i\ep'\,\bk'\,a';
i\ep'+i\omega\,\bk'\,b',i\ep\,\bk\, b\big)\\
&\quad =\Gamma_{0}\big(i\ep'\,\bk'\,a',i\ep+i\omega\,\bk\,a;i\ep\,\bk\, b,
i\ep'+i\omega\,\bk'\,b'\big)\,,
\eal
the irreducible vertex in the particle-hole channel with transferred frequency $\omega$. Since
\bealn
&\Gamma_{0}\big(i\ep+i\omega\,\bk\,a,i\ep'\,\bk'\,a';
i\ep'+i\omega\,\bk'\,b',i\ep\,\bk\, b\big)^*\\
&\quad =\Gamma_{0}\big(-i\ep\,\bk\,b,-i\ep'-i\omega\,\bk'\,b';
-i\ep'\,\bk'\,a',-i\ep-i\omega\,\bk\,a
\big)\,,
\eal
and $G_{ba}(i\ep,\bk)^*=G_{ab}(-i\ep,\bk)$, one can readily show that 
$K_L$ in \eqn{K_L} is indeed real. We observe that $F$ in \eqn{F-Gamma} is 
odd under $(i\ep,\bk,ab)\leftrightarrow (i\ep,\bk',a'b')$, unlike what claimed in \cite{Ishikawa-1, Ishikawa-2}, which compensates the change of sign of the 
antisymmetric tensor $\ep_{ij}$ under $i \leftrightarrow j$. Therefore, 
$K_L$ may well be finite in principle. Nonetheless, one may still argue that 
$F$ could vanish. Indeed, the irreducibility in the particle-hole channel suggests that $\Gamma_0$ depends on 
$i\omega$ only through the frequency carried by the particle-particle channel, as can be verified by inspection of few orders in the skeleton expansion. That 
frequency remains invariant if in the top panel of Fig.~\ref{fig1} we change 
the frequency of the upper Green's function from $i\ep+i\omega$ to $i\ep$, and that of the lower one from $i\ep$ to $i\ep+i\omega$, thus $\omega\to-\omega$ in the particle-hole channel. If that is true, $\Gamma_0$ is even in $\omega$, and 
since it must also be smooth, then its derivative at $\omega=0$ has to vanish, as also argued by  
Ref.~\cite{Gabi-PRB2013} on the basis of the skeleton expansion. However, once perturbation theory breaks down, one cannot exclude that 
the full series develops odd contributions, and thus that   
$K_L$ becomes non zero. \\

\noindent
Further insight in that direction can be gained through the Streda 
formula~\cite{Streda_1982,Streda_1983}
\beal
\fract{\partial\rho}{\partial B}_{\big|B=0} &= \fract{e}{2\pi c}\;\nu_\text{H}\,,
\label{Streda}
\eal
where $\rho$ is the electron density. The spin-Hall conductance is defined similarly provided $\rho$ is replaced by the spin density~\cite{Yang-PRB2006}. 
The electron density at $T=0$ can be calculated through~\cite{Jan}
\beal
\rho &= \fract{n}{2} + \int \fract{d\ep\, d\bk}{(2\pi)^3}\;
\Tr\Big(G(i\ep,\bk)\Big) \\
&= \fract{n}{2} - \int \fract{d\ep\, d\bk}{(2\pi)^3}\;
\fract{\partial \ln\text{det}\,G(i\ep,\bk)}{\partial i\ep} + I_L\\
&= \fract{n}{2} - \int \fract{d\ep\, d\bk}{(2\pi)^3}\;
\fract{\partial \ln\text{det}\,G_*(i\ep,\bk)}{\partial i\ep} + I_L\\
&= \int \fract{d\bk}{(2\pi)^2}\;\theta\big(-\ep_\alpha(0,\bk)\big)+I_L
\,,
\label{density-def}
\eal
where $n$ is the number of bands, including spin, we used the fact that 
\bealn
\ln\text{det}\,G(i\ep,\bk)=\ln\text{det}\,G_*(i\ep,\bk)
+\ln\text{det}\,Z(\ep,\bk)\,,
\eal
and $\ln\text{det}\,Z(\ep,\bk)$ gives no contribution to the integral since its 
derivative is odd, and, finally,  
\bealn
I_L &= \int \fract{d\ep\, d\bk}{(2\pi)^3}\;
\Tr\bigg(G(i\ep,\bk)\;\fract{\partial \Sigma(i\ep,\bk)}{\partial i\ep}\bigg)\\
&=\fract{1}{\pi}\,\int \fract{d\bk}{(2\pi)^2}\, \Ima\,\mathcal{I}(i0^+,\bk)\,.
\eal
The function $\mathcal{I}(i\ep,\bk)=\mathcal{I}(-i\ep,\bk)^*$ is defined as~\cite{Jan}
\bealn
\mathcal{I}(i\ep,\bk)&= \Phi(i\ep,\bk)-\Tr\Big(\Sigma(i\ep,\bk)\,G(i\ep,\bk)\Big)\,,
\eal
having written the Luttinger-Ward functional, which can be constructed  
fully non-perturbatively~\cite{Potthoff-2006}, as 
\bealn
\Phi[G] &= T\sum_\ell\,\int \fract{d\bk}{(2\pi)^2}\; \esp{i\ep_\ell 0^+}\;
\Phi(i\ep_\ell,\bk)\,.
\eal
In the perturbative regime, $\Ima\,\mathcal{I}(i\ep,\bk)\sim\ep$ for small $\ep$ 
and thus $I_L=0$, in which case \eqn{density-def} reduces to the well known statement of Luttinger's theorem~\cite{Luttinger}. However, when perturbation theory breaks down, $I_L$ is generally nonzero~\cite{Jan}. \\
In presence of a magnetic field $B$, the derivative of the first term in \eqn{density-def} with respect to $B$ and calculated at $B=0$ yields the 
Streda formula for non-interacting electrons described by the quasiparticle 
Hamiltonian $H_*(0,\bk)$, which is just the TKNN expression \eqn{final-W}. 
It follows that 
\beal
\fract{\partial I_L}{\partial B}_{\big| B=0} &\equiv \fract{e}{2\pi c}\;
K_L\,,
\label{I_L vs K_L}
\eal  
namely that the winding number \eqn{W} may not correspond to the quantised Hall conductance when perturbation theory breaks down and $I_L\not=0$, in accordance, e.g., with the results of \cite{Slage-PRB2015,He-PRB2016-II}.

\section{Properties of the winding number}
\label{Properties of the winding number}

Using the property of the antisymmetric tensor, the equivalence 
$G(i\ep,\bk)^\dagger=G(-i\ep,\bk)$, and the fact that one of the derivatives 
in the winding number \eqn{W} 
is $\partial_\ep=-\partial_{-\ep}$, one finds that
\bw
\bealn
W(G)^* &= \fract{1}{24\pi^2}\int \!d\ep\,d\bk\,\ep_{\mu\nu\rho}\,\Tr\Big(G(-i\ep,\bk)\,\partial_\rho G(-i\ep,\bk)^{-1}\, G(-i\ep,\bk)\,\partial_\nu G(-i\ep,\bk)^{-1}\,
G(-i\ep,\bk)\,\partial_\mu G(-i\ep,\bk)^{-1}\Big)\\
&= -\fract{1}{24\pi^2}\int \!d\ep\,d\bk\,\ep_{\mu\nu\rho}\,\Tr\Big(G(i\ep,\bk)\,\partial_\rho G(i\ep,\bk)^{-1}\, G(i\ep,\bk)\,\partial_\nu G(i\ep,\bk)^{-1}\,
G(i\ep,\bk)\,\partial_\mu G(i\ep,\bk)^{-1}\Big)\\
&= \fract{1}{24\pi^2}\int \!d\ep\,d\bk\,\ep_{\rho\nu\mu}
\,\Tr\Big(G(i\ep,\bk)\,\partial_\rho G(i\ep,\bk)^{-1}\,
G(i\ep,\bk)\,\partial_\nu G(i\ep,\bk)^{-1}\,
G(i\ep,\bk)\,\partial_\mu G(i\ep,\bk)^{-1}\Big)= W(G)\,,
\eal
\ew
thus that $W(G)$ is real, as expected. Now suppose that the function 
$G(i\ep,\bk)$, beside satisfying \eqn{G-property}, which guarantees that $W(G)$ is real, is 
also hermitian. It follows that $G(i\ep,\bk)=G(i\ep,\bk)^\dagger=G(-i\ep,\bk)$, thus 
$G(i\ep,\bk)$ is even in $\ep$, hence its derivative is odd. As a consequence, the winding number 
trivially vanishes being an integral over $\ep\in[-\infty,\infty]$ of an odd function. This is precisely the reason why $W(Z)=0$, since the quasiparticle residue matrix $Z(\ep,\bk)$, Eq.~(3) of the main text, 
is hermitian and even in $\ep$.    \\

\subsection{Proof of $W(G_1\,G_2)=W(G_1)+W(G_2)$}
We shortly write 
\bealn
W(G) &= \fract{1}{24\pi^2}\,\int d\bx \,\ep_{\mu\nu\rho}
\,\Tr\Big(G(\bx)\,\partial_\mu G(\bx)^{-1}
\\
&\qquad \qquad \qquad \qquad G(\bx)\,\partial_\nu G(\bx)^{-1}\;
G(\bx)\,\partial_\rho G(\bx)^{-1}\Big)\,,
\eal
where $\bx=\big(x_0,x_1,x_2\big)$ is a three component vector. Assume that 
$G(\bx)=G_1(\bx)\,G_2(\bx)$, then, since
\bealn
G\,\partial_\mu G &= 
G_1\, G_2 \;\partial_\mu\left( G_2^{-1}\, G_1^{-1} \right) \\
&= G_1 \left( G_2\; \partial_\mu G_2^{-1} + \partial_\mu G_1^{-1}\; G_1 \right) G_1^{-1} \, ,
\eal
the winding number of the product can be written as, defining, for simplicity,  $G\equiv G(\bx)$, 
\bw 
\beal
W(G_1 G_2) &= \fract{1}{24\pi^2}\,\int d\bx \,\ep_{\mu\nu\rho}\,\Tr\bigg[
\left( G_2\; \partial_\mu G_2^{-1} + \partial_\mu G_1^{-1}\; G_1 \right) \left( G_2\; \partial_\nu G_2^{-1} + \partial_\nu G_1^{-1}\; G_1 \right) \left( G_2\; \partial_\rho G_2^{-1} + \partial_\rho G_1^{-1}\; G_1 \right)\bigg]\\
&= W(G_1) + W(G_2) \\
&\quad + \fract{3}{24\pi^2}\,\int d\bx \,\ep_{\mu\nu\rho}\, \Tr\bigg[
G_2\;\partial_\mu G_2^{-1}\;
G_2\;\partial_\nu G_2^{-1}\;\partial_\rho G_1^{-1}\;
G_1\, + \, \partial_\mu G_1^{-1}\;G_1\;\partial_\nu G_1^{-1}\;G_1\;G_2\;\partial_\rho G_2^{-1}\bigg] \,.
\eal
\ew
Since $G \,\partial_\mu G^{-1} = - \partial_\mu G\; G^{-1}$ and 
$\epsilon_{\mu \nu \rho} \,  \Big(\partial_\mu \partial_\nu G\Big) = 0 $ if $G$ is smooth, 
we find that 
\bw
\beal
I&\equiv \ep_{\mu\nu\rho}\,\Tr\bigg[ G_2\;\partial_\mu G_2^{-1}\;
G_2\;\partial_\nu G_2^{-1}\;\partial_\rho G_1^{-1}\;
G_1\, + \, \partial_\mu G_1^{-1}\;G_1\;\partial_\nu G_1^{-1}\;G_1\;G_2\;\partial_\rho G_2^{-1}\bigg] \\
& = - \ep_{\mu\nu\rho}\,\Tr\bigg[\partial_\mu G_2\;\partial_\nu G_2^{-1}\;\partial_\rho G_1^{-1}\;
G_1\, + \, \partial_\mu G_1^{-1}\;\partial_\nu G_1\;G_2\;\partial_\rho G_2^{-1}\bigg] \\
&= - \ep_{\mu\nu\rho}\,\Tr\bigg[\partial_\mu \Big(G_2\;\partial_\nu G_2^{-1}\Big)\;\partial_\rho G_1^{-1}\;
G_1\, + \partial_\nu \Big(\partial_\mu G_1^{-1}\; G_1\Big)\;G_2\;\partial_\rho G_2^{-1}\bigg] \, .
\eal
\ew
Integrating by part the first term, and then switching the indexes $\mu \leftrightarrow \nu $ in the first term and $\mu \leftrightarrow \rho $ in the second one, we obtain
\bw
\beal
I &= - \ep_{\mu\nu\rho}\,\Tr\bigg[- G_2\;\partial_\nu G_2^{-1}\;\partial_\mu \Big( \partial_\rho G_1^{-1}\;
G_1 \Big) + \, G_2\;\partial_\rho G_2^{-1} \; \partial_\nu \Big(\partial_\mu G_1^{-1}\; G_1\Big)\bigg] \\
&= - \ep_{\mu\nu\rho}\,\Tr\bigg[G_2\;\partial_\mu G_2^{-1}\;\partial_\nu \Big( \partial_\rho G_1^{-1}\;
G_1 \, \big) - G_2\;\partial_\mu G_2^{-1} \; \partial_\nu \Big(\partial_\rho G_1^{-1}\; G_1\Big)\bigg] 
= 0 \, ,
\eal
\ew
which proves that 
\beal
W\big(G_1\,G_2\big) = W(G_1) + W(G_2)\;.
\eal

\bibliographystyle{apsrev4-2}

\end{document}